\documentclass[twocolumn,superscriptaddress,amsmath,amssymb,pra,longbibliography,reprint]{revtex4-2}
\usepackage{listings}
\usepackage{amsmath}
\usepackage{amssymb}
\usepackage{amsxtra}
\usepackage{physics}
\usepackage{mathrsfs}
\usepackage{booktabs}
\usepackage{gensymb}
\usepackage{graphicx}
\usepackage{tikz}
\usepackage{hyperref}
\usepackage{lipsum}
\usepackage{comment}

\usepackage[TS1,T1]{fontenc}
\usepackage{textcomp}

\newcommand{\ui}{\mathrm{i}}
\usepackage{mathtools}

\usepackage{makecell}

\begin{document}

\title{Orbital angular momentum of spatiotemporal vortices: \\ a ray-mechanical analogy}

%\title{STOVs orbital angular momentum: a simple mechanical analogy}

\author{Sophie Vo}
\affiliation{The Institute of Optics, University of Rochester, Rochester, NY 14627, USA}

\author{Konstantin Y. Bliokh}
\affiliation{Donostia International Physics Center (DIPC), Donostia-San Sebasti\'an 20018, Spain}
\affiliation{IKERBASQUE, Basque Foundation for Science, Bilbao 48009, Spain}
\affiliation{Centre of Excellence ENSEMBLE3 Sp.~z o.o., 01-919 Warsaw, Poland}

\author{Miguel A. Alonso}
\affiliation{Aix Marseille Univ, CNRS, Centrale Med, Institut Fresnel, UMR 7249, Marseille Cedex 20, 13397, France}
\affiliation{The Institute of Optics, University of Rochester, Rochester, NY 14627, USA}

\begin{abstract}
Spatiotemporal vortex pulses (STVPs) are wavepackets that carry transverse orbital angular momentum (OAM), whose proper quantification has been the subject of recent debate. In this work, we introduce a simplified mechanical model of STVPs, consisting of a loop of non-interacting point particles traveling at a uniform constant speed but at slightly different angles. We examine different initial conditions for the particle loop, including configurations that are elliptic in space at a given time and configurations that are elliptic in spacetime at a fixed propagation distance. Furthermore, employing a non-uniform mass distribution allows the particle loop to mimic the STVP not only in configuration space but also in momentum space. Remarkably, when supplemented by a semiclassical vorticity quantization condition, our mechanical model exactly reproduces different wave-based OAM results previously reported for paraxial STVPs.
\end{abstract}

\maketitle

%%%%%%%%%%%%%%%%%%%%%%%%%%%%%%%%%%%%%%
\section{Introduction}
%%%%%%%%%%%%%%%%%%%%%%%%%%%%%%%%%%%%%%

Wave vortices are fundamental structures in inhomogeneous wave fields of any nature, both classical \cite{Soskin2001, Dennis2009} and quantum \cite{Bliokh2017PR}. These are characterized by phase singularities, i.e., points of vanishing intensity with dislocations of the wavefront \cite{Nye1974}. For localized wave beams or packets, wave vortices are also commonly associated with the {\it orbital angular momentum (OAM)} carried by the wave \cite{Allen_book, Andrews_book, Bliokh2015PR, Bliokh2025CP}.    

Until fairly recently, most studies of wave vortices focused on monochromatic waves and stationary vortices. In particular, axially symmetric vortex beams support vortices whose normalized OAM is directed along the beam axis (i.e., along the mean wave momentum) and is unambiguously quantified (for scalar waves) by the integer vortex strength $\ell$ \cite{Allen1992PRA, Ceperley1992AJP, Allen_book, Bliokh2015PR}. Notably, this {\it longitudinal} OAM is purely {\it intrinsic}, i.e., invariant with respect to the choice of coordinate origin \cite{Berry1998, Bliokh2015PR}.  

In the more general case of non-monochromatic waves, vortices can be non-stationary, as already recognized in the pioneering work of Nye and Berry \cite{Nye1974}. Recently, such {\it spatiotemporal vortices} have attracted considerable attention, motivating theoretical \cite{Sukhorukov2005, Bliokh2012PRA, hancock2021mode, bliokh2023orbital, Porras2023OL, porras2023transverse, vo2024closed1, Wang2021O, bekshaev2024spatiotemporal} and experimental \cite{Jhajj2016, Chong2020NP, Hancock2019O, Zang2022NP, Martin-Hernandez2025NP, Zhang2023NC, Ge2023PRL, Che2024PRL} studies across optical, acoustic, quantum, and water-wave systems. 

Localized spatiotemporal vortices, referred to as {\it spatiotemporal vortex pulses (STVPs)}, carry OAM that is not aligned with the mean momentum \cite{Bliokh2015PR, Bliokh2012PRA}. In the limiting case of the vortex moving orthogonally to the wavefront dislocation line, the OAM becomes purely {\it transverse} \cite{Chong2020NP, hancock2021mode, bliokh2023orbital, bekshaev2024spatiotemporal, porras2023transverse}. 

Remarkably, theoretical calculations of such transverse OAM encountered significant controversies and debated the magnitude of the transverse OAM carried by STVPs \cite{hancock2021mode, bliokh2023orbital, porras2023transverse, porras2024clarification, bekshaev2024spatiotemporal, Tripathi2025OE, Gadeyne2025}. The main origin for these discrepancies lies in the strong dependence of transverse OAM on the choice of coordinate origin, which implies an essential {\it extrinsic} OAM part \cite{Bliokh2015PR, bliokh2023orbital}.  
%Recent works \cite{hancock2021mode, bliokh2023orbital, porras2023transverse, porras2024clarification, bekshaev2024spatiotemporal, Tripathi2025OE, Gadeyne2025} have debated the magnitude of the transverse OAM carried by STVPs. 
Differing results have been obtained for the OAM and its intrinsic and extrinsic parts, as summarized in Table~\ref{tab:priorart}. It has been suggested that these discrepancies originate from the choice of STVP's shape and reference point.

%There has been a discussion of the different shapes that can be used for the calculation of the properties of spatiotemporal optical vortices (STOVs). Some previous studies assume STOVs that are elliptical in space, while others consider that STOVs are elliptical at a given cross section and time, that is, in spacetime. %Here, we use a ray-optics based formalism allowing to understand that the shape of a STOV depends on which framework is used to describe the propagation of this type of wavepacket. 
%The debate centers around the amount of orbital angular momentum carried by STOVs: while all studies separate this amount into an intrinsic and an extrinsic part, different results are obtained for the respective intrinsic and extrinsic contributions, as shown in Table~\ref{tab:priorart}. Some researchers suspect that the discrepancy in results come from differences in shape. Some others presume that the divergence originates from the chosen points of reference.

%Here we present a derivation using a simplistic picture based on free particles forming a loop that evolves like a STOV. We find that, while extremely simplified, this model lets us understand the differences in an intuitive way.

%Tables~\ref{tab:Bliokh1}  to~\ref{tab:Berkshaev1} present OAM expressions of free-space propagating STOVs described in previous works. All those works share in common that the OAM expression involves the topological charge of the STOV and its ellipticity or eccentricity  either in 3D space or in spacetime.

%TTTTTTTTTTTTTTTTTTTTTTTTTTTTTTTTTTTTTTTTTTTTTTTTTTTTTTTTTTTTTTT
\begin{table}[t]
\begin{center}
    \caption{Expressions for the transverse OAM of STVPs in prior art. Here, $\ell$ is the STVP's topological charge and $\gamma$ is the ratio of the longitudinal and transverse dimensions of the STVP.}
%\hspace*{0.8cm}
\renewcommand{\arraystretch}{2}
\begin{tabular}{c|c} 
\hline \hline 
\textbf{Reference} & \textbf{OAM} \\ 
%[0.5ex] 
\hline
{Bliokh \cite{bliokh2023orbital}, $xz$ framework} & $ L_y = \dfrac{\gamma}{2}\, \ell$ \\
{and}  & $ L^{\rm INT}_y \!=  \dfrac{\gamma+ \gamma^{-1}}{2} \ell$ \\
Bekshaev \cite{bekshaev2024spatiotemporal}, $xt$ framework & $ L^{\rm EXT}_y = -\dfrac{1}{2\gamma}\, \ell$ \\
 \hline
{Hancock {\it et al.} \cite{hancock2021mode}, $xt$ framework} & $ L_y = \dfrac{\gamma}{2} \ell$ \\
 \hline
 & $ L^{\rm INT}_y =\dfrac{\gamma}{2}\, \ell$ \\
  {Porras \cite{porras2023transverse}, $xt$ framework}
  & $L_y^{\rm EXT} = -\dfrac{\gamma}{2}\, \ell$ \\
  & $L_y = 0$ \\
\hline
% Bekshaev's OAM - spacetime framework \cite{bekshaev2024spatiotemporal} &  $\langle L_y\rangle / W$ &  $\frac{1}{2} \frac{1}{\omega_0} \gamma \ell$  \\
%  \quad &$\langle L_y^{\rm int}\rangle / W$ &  $\frac{1}{2} \frac{1}{\omega_0} (\gamma + \frac{1}{\gamma}) \ell$  \\
%\quad & $\langle L_y^{\rm ext}\rangle / W$ &   $-\frac{1}{2} \frac{1}{\omega_0} \frac{1}{\gamma}$ \ell$ \\
\hline
\end{tabular}
%\hspace*{-1cm}
\label{tab:priorart}
\end{center}
\end{table}
%TTTTTTTTTTTTTTTTTTTTTTTTTTTTTTTTTTTTTTTTTTTTTTTTTTTTTTTTTTTTTTT

Here we re-examine the transverse OAM of STVPs, using an approach that differs substantially from the methods used in previous studies.
Instead of relying on a {\it wave}-based analysis, we put forward a simplified {\it mechanical} model similar to the geometrical ray approach.  
Despite its simplicity, this model reproduces the main results of prior wave calculations and provides an intuitive framework that sheds light on the origins of the previously-reported differences.

%%%%%%%%%%%%%%%%%%%%%%%%%%%%%%%%%%%%%%
\section{Model of a STVP as a particle loop with uniform mass density}
\label{sec:II}
%%%%%%%%%%%%%%%%%%%%%%%%%%%%%%%%%%%%%%

%%%%%%%%%%%%%%%%%%%%%%%%%%%%%%%%%%%%%%
\subsection{$xz$ and $xt$ frameworks for elliptical STVPs}
\label{subsection:frameworks}
%%%%%%%%%%%%%%%%%%%%%%%%%%%%%%%%%%%%%%

\begin{figure}[h]
     \centering
    \includegraphics[width=\linewidth]{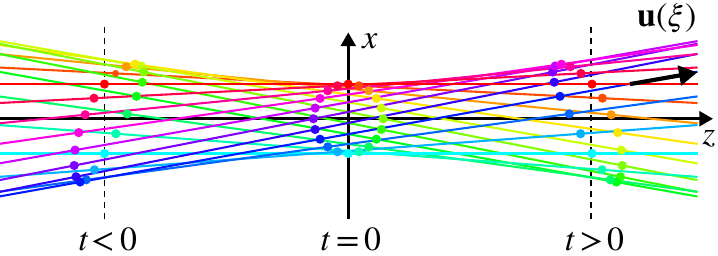}
     \caption{ Rays corresponding to trajectories of particles for modeling the evolution of STVPs \cite{vo2024closed1}. In all figures, hue colors are used to identify the value of $\xi \in [0,2\pi)$ for the corresponding ray.}
      \label{fig:rays}
\end{figure}

STVPs are wavepackets that can be seen as propagating in space and evolving in time. 
It was recently shown that the diffractive deformation of STVPs with propagation can be understood by considering a non-interacting set of particles, each running with the same speed $c$ along a rectilinear trajectory referred to here as a {\it ray} 
\cite{vo2024closed1}, as illustrated in Fig.~\ref{fig:rays}. 
Each ray in the set corresponds to a value of a periodic parameter $\xi \in [0,2\pi)$. 
Assuming a STVP propagating in the $(x,z)$ plane along the $z$-axis, the coordinates of each particle can be written in the following form:
\begin{align}
{\bf r}(\xi,t)=
%\begin{pmatrix}
%X(\xi,t) \\
%Z(\xi,t) 
%\end{pmatrix}=
\begin{pmatrix}
X_0(\xi) \\
Z_0(\xi) 
\end{pmatrix}
+ [t-t_0(\xi)] \,c\,{\bf u}(\xi)\,,
\label{eq:generalrays}
\end{align}
where ${\bf u}(\xi)$ is a unit vector defining the direction of each ray:
\begin{align}
{\bf u}(\xi)=
%\left(\begin{array}{c}u_x(\xi)\\u_z(\xi)\end{array}\right)=
\frac{1}{\sqrt{1+\Delta_x^2 \cos^2{\xi}}}\!\left(\begin{array}{c}\Delta_x \cos{\xi}\\1\end{array}\right)
\!\simeq\! \left(\begin{array}{c}\Delta_x \cos{\xi}\\1\end{array}\right)\!.
\label{eq:px}
\end{align}
Here, $\Delta_x$ is the maximal magnitude of the ray slope with respect to the $z$ axis, and the linear approximation in $\Delta_x \ll 1$ in the second step corresponds to the paraxial approximation. 

%FFFFFFFFFFFFFFFFFFFFFFFFFFFFFFFFFFFFFFFFFFFFFFFFFFFFFFFFFFFFFF
%FFFFFFFFFFFFFFFFFFFFFFFFFFFFFFFFFFFFFFFFFFFFFFFFFFFFFFFFFFFFFF

\begin{figure}[t]
\centering
\includegraphics[width=\linewidth]{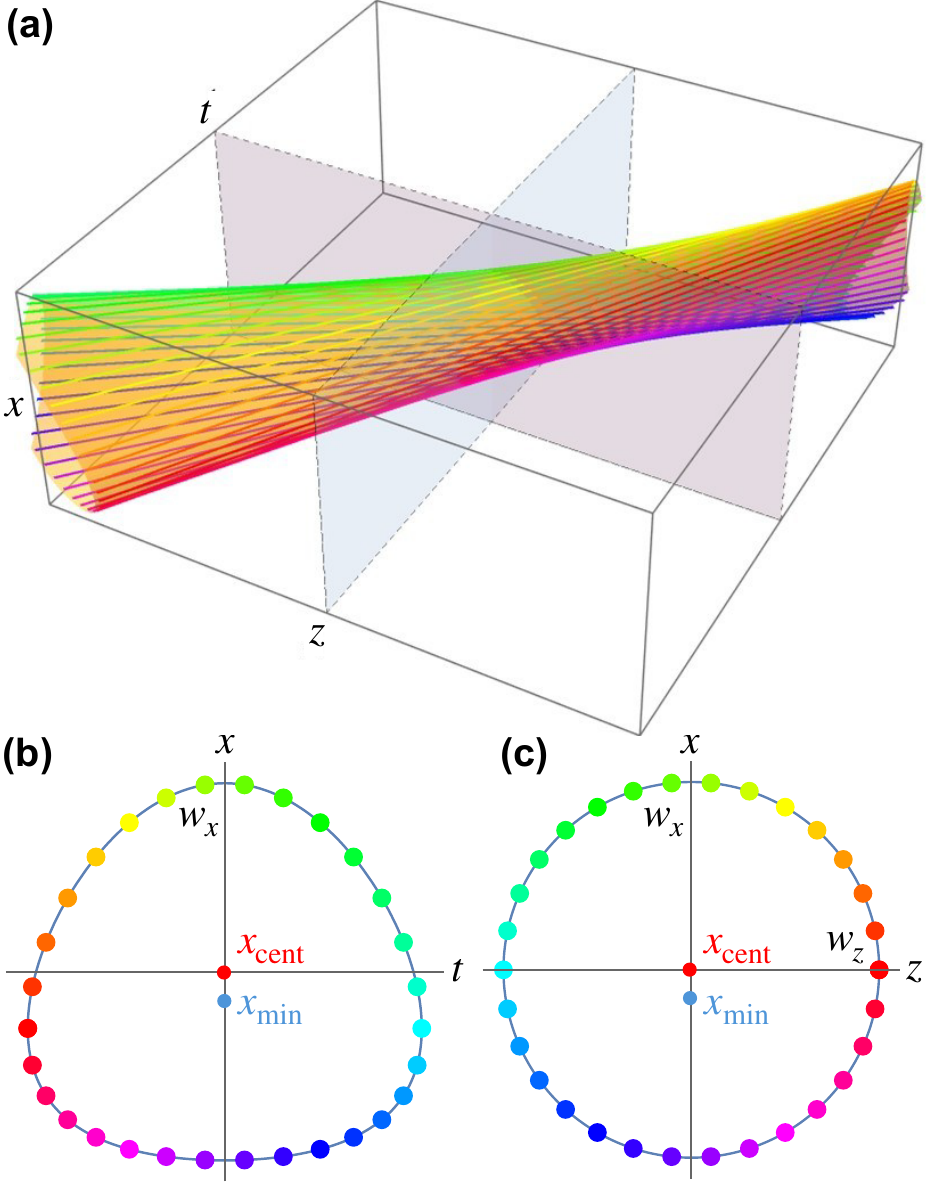}
\caption{Ray-optics modeling of $xz$ STVPs: (a) Ray bundle in space and time for $w_x=w_z$ and $\Delta_x=0.3$; % a reference frame $(z,ct,x)$; 
(b,c) Cross-sections at $z=0$ (b), and $t=0$ (c), where the reference points $x_{\rm cent}$ and $x_{\rm min}$ are also indicated. }
\label{fig:Teq0_ZX}
\end{figure}

The functions $X_0$, $Z_0$ and $t_0$ in Eq.~(\ref{eq:generalrays}) determine the initial conditions (and by consequence the general shape) of the set of particles. Note that one of them is redundant, but using all three simplifies the analysis that follows.  We now consider two particular choices:

%%%%%%%%%%%%%%%%%%%%%%%%%%%%%%%%%%%%%%
%\red{[MA: I wonder if the subsection titles A and B can be removed, and just made in-line boldface titles for each paragraph. Could we add a more general subsection title for this part?]}
%\subsection{Elliptical particle loop over the $xz$ plane at $t=0$}
%\label{subsection:ZX-attempt1}
%%%%%%%%%%%%%%%%%%%%%%%%%%%%%%%%%%%%%%
{\bf {\it xz} STVPs:} In the first framework, we consider the following initial conditions:
\begin{align}
X_0(\xi)&= w_x \sin{\xi}\,,\quad
Z_0(\xi)= w_z \cos{\xi}\,,\quad
t_0(\xi)=0\,.
\label{Initial_1}
\end{align}
These conditions are such that the particles trace over the $zx$ plane at $t=0$ an ellipse of semi-axes $w_z$ and $w_x$ centered at the origin, as shown in Fig.~\ref{fig:Teq0_ZX}(c). 
The substitution of initial conditions \eqref{Initial_1} into Eq.~(\ref{eq:generalrays}) 
can be used to define a parameterized surface in spacetime, as illustrated in Fig.~\ref{fig:Teq0_ZX}(a).
Note from Fig.~\ref{fig:Teq0_ZX}(b) that the cross-section of this surface by the $xt$-plane at $z=0$ is not elliptic. The propagation of the particles is also illustrated in Supplementary Movie 1.

%, where a subset of the rays are also shown. 
%a reference frame $(z,ct,x)$. Figure~\ref{fig:Teq0_ZX}a shows a representation of that three-dimensional surface, for the case $w_x=w_z$, where a set of rays corresponding to the description of this surface by varying  the parameter $\xi$ between 0 and $2 \pi$ has been added.   

%FFFFFFFFFFFFFFFFFFFFFFFFFFFFFFFFFFFFFFFFFFFFFFFFFFFFFFFFFFFFFF
%FFFFFFFFFFFFFFFFFFFFFFFFFFFFFFFFFFFFFFFFFFFFFFFFFFFFFFFFFFFFFF

%%%%%%%%%%%%%%%%%%%%%%%%%%%%%%%%%%%%%%
%\subsection{Elliptical particle loop over the $xt$ plane at $z=0$}
%\label{subsection:TX-attempt1}
%%%%%%%%%%%%%%%%%%%%%%%%%%%%%%%%%%%%%%

\begin{figure}[t]
\centering
\includegraphics[width=\linewidth]{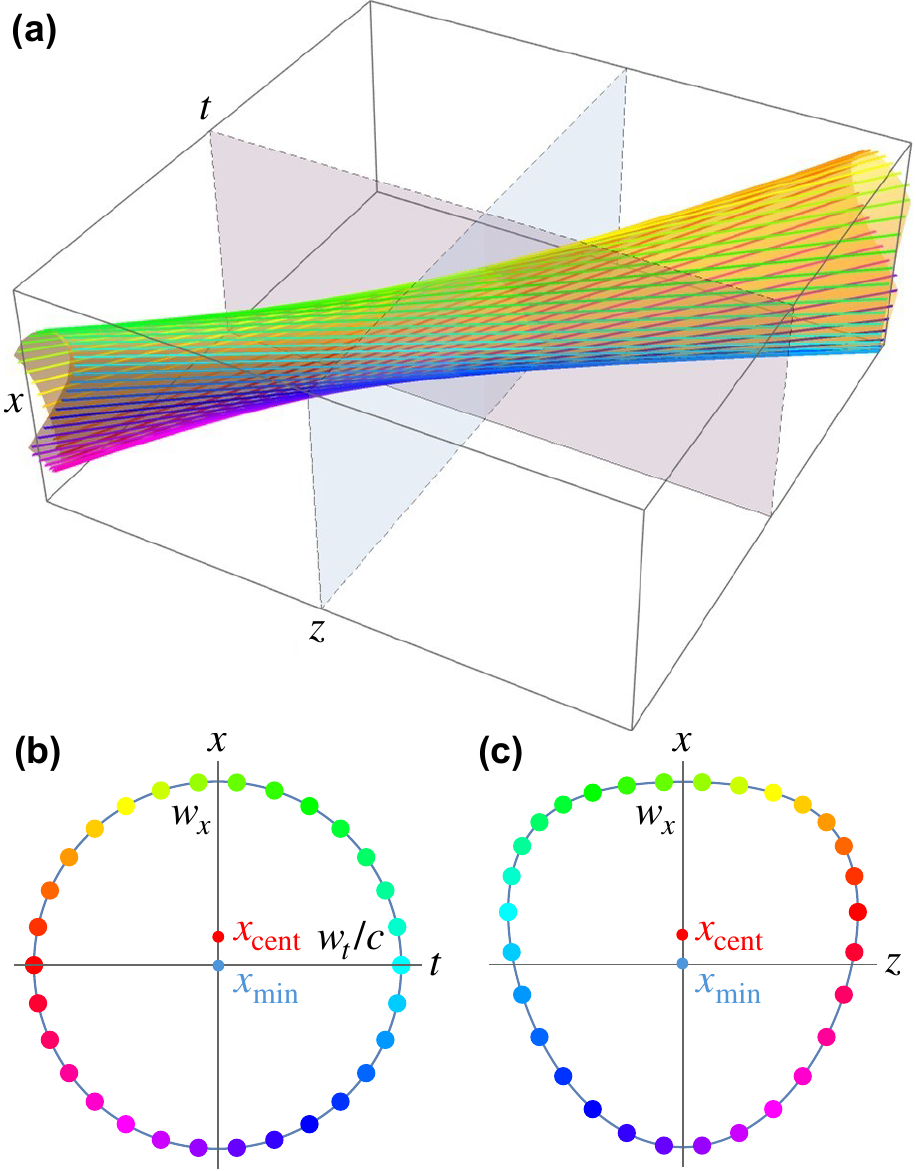}
\caption{ Ray-optics modeling of $xt$ STVPs: (a) Ray bundle in space and time for $w_x=w_t$ and $\Delta_x=0.3$; (b,c) Cross-sections at $z=0$ (b), and $t=0$ (c), where the reference points $x_{\rm cent}$ and $x_{\rm min}$ are also indicated.}
\label{fig:Zeq0_TX}
\end{figure}

{\bf {\it xt} STVPs:} In the second framework, we choose instead the initial conditions
\begin{align}
X_0(\xi)&= w_x \sin{\xi}\,,~~
Z_0(\xi)=0\,,~~
t_0(\xi)= - \frac{w_t}{c} \cos{\xi}\,.
\label{Initial_2}
\end{align}
In contrast with the previous case, these conditions guarantee that the particles trace at $z=0$ an ellipse of semi-axes $w_x$ and $w_t/c$ centered at the origin over the $xt$-plane, as illustrated in  \ref{fig:Zeq0_TX}(c). 
Similarly to the previous case, the parameterized surface defined by Eq.~(\ref{eq:generalrays}) 
with initial conditions \eqref{Initial_2} can be plotted in spacetime, as shown in Fig.~\ref{fig:Zeq0_TX}(a).
We see from Fig.~\ref{fig:Zeq0_TX}(c) that the cross-section of this spacetime surface by the $xz$-plane at $t=0$ has a non-elliptical shape. The propagation of the particle loop is  illustrated in Supplementary Movie 2.

We note that the simple ray-optical profiles in Figs.~\ref{fig:Teq0_ZX}(b,c) and~\ref{fig:Zeq0_TX}(b,c) resemble the shapes of the wave-based calculations by Porras \cite{porras2024clarification}, which used elliptical intensity distributions either in the $(x,z)$ plane or in the $(x,t)$ plane. %In what follows, we will refer to these two frameworks as the $xz$ and $xt$ frameworks, respectively.

%FFFFFFFFFFFFFFFFFFFFFFFFFFFFFFFFFFFFFFFFFFFFFFFFFFFFFFFFFFFFFF
%FFFFFFFFFFFFFFFFFFFFFFFFFFFFFFFFFFFFFFFFFFFFFFFFFFFFFFFFFFFFFF
      
%%%%%%%%%%%%%%%%%%%%%%%%%%%%%%%%%%%%%%
\subsection{Orbital angular momentum}
%%%%%%%%%%%%%%%%%%%%%%%%%%%%%%%%%%%%%%
We now compare these two frameworks and calculate the OAM for the corresponding particle loops. We start by assuming classical particles of the same mass propagating at the same speed, and hence carrying linear momentum density (in terms of the parameter $\xi$) 
\begin{align}
{\bf p}(\xi) = \frac{mc}{2\pi} {\bf u}(\xi), 
\label{eq:plinear}
\end{align}
where $m$ is the total mass of the set of particles. Then, the OAM computed with respect to a chosen point ${\bf r}_0 = (0,x_0)$ at the $z=0$ plane is defined by 
%\red{[MA: Now that ${\bf p}$ is defined, I wonder if we should write this next equation in terms of momentum, which is more familiar and then we do not need to rewrite it in the next subsection?]}
%
\begin{align}
\langle {\bf L}\rangle(x_0)=
%\frac{mc}{2 \pi} \int\limits_{0}^{2 \pi}[{\bf r}(\xi,t) - {\bf r}_0] \times {\bf u}(\xi)\,{\rm d}\xi\,.
\int\limits_{0}^{2 \pi}[{\bf r}(\xi,t) - {\bf r}_0] \times {\bf p}(\xi)\,{\rm d}\xi\,.
\label{eq:OAM}
\end{align}
%
%where $m$ is the total mass of the set of particles and $c$ is their speed. In this section, we make the assumption that the mass density is uniform in $\xi$.
%, equal to $\frac{mc}{2 \pi}$. 
%In other words, all infinitesimal subsets of particles corresponding to a given differential increment ${\rm d}\xi$ contain the same mass. 
It is easy to see from Eqs.~\eqref{eq:generalrays} and \eqref{eq:plinear} that the OAM \eqref{eq:OAM} is independent of $t$ (so it is sufficient to set $t=0$), 
it is directed along the $y$ axis (normal to the $zx$-plane), and its magnitude depends on the choice of the reference position $x_0$. 

Substituting Eqs.~\eqref{eq:generalrays}--\eqref{Initial_2} into Eq.~\eqref{eq:OAM}, we find that the OAM of an STVP in the $xz$ and $xt$ frameworks are given (in the paraxial regime) by
\begin{align}
\langle {L_y^{(xz)}}\rangle(x_0) & = mc\left( x_0 + \frac{\Delta_x w_z}{2} \right), \nonumber \\
\langle {L_y^{(xt)}}\rangle(x_0) & = mc\, x_0\,.
\label{eq:OAM_1}
\end{align}

In order to calculate an ``intrinsic'' OAM, one has to choose an appropriate $x_0$ not simply as the origin, but according to some meaningful criterion in this mechanical analogy. We consider the following two options:
\begin{itemize}
\item %A first option is to compute it at 
the transverse centroid of the particles at $t=0$, defined by
\begin{align}
x_{\rm cent}=\frac{1}{2 \pi} \int\limits_{0}^{2 \pi}x(\xi,t=0)\, {\rm d}\xi\,;
\label{eq:xcent}
\end{align}
\item %A second option is to compute the OAM at 
the point (0,$x_{\rm min}$) that minimizes the square of the OAM \cite{alonso2000uncertainty}. More specifically, $x_{\rm min}$ is the value of $x_0$ that minimizes the quantity
\begin{align}
\langle L^2\rangle(x_0)=\frac{m^2c^2}{2 \pi} \int\limits_{0}^{2 \pi} \left|[{\bf r}(\xi,t=0) - {\bf r}_0] \times {\bf u}(\xi) \right|^2 {\rm d}\xi\,.
\label{eq:OAM-min}
\end{align}
\end{itemize}

The exact expressions (valid in the nonparaxial regime) for the different centroids and the corresponding OAM are summarized in the tables in the Supplementary Document A. 
%Appendix~\ref{AppendixA}. 
In the paraxial regime, linear in $\Delta_x \ll 1$, we obtain for the $xz$ and $xt$ STVPs the centroids and corresponding OAM \eqref{eq:OAM_1} summarized in Table~\ref{tab:results-paraxial}. Notably, the OAM values evaluated either at $x_{\rm cent}$ or $x_{\rm min}$ are similar in the paraxial regime for both types of STVP: $xz$-elliptical and $xt$-elliptical. 
Thus, for this simple particle model, shape itself is not the source of discrepancy between different results when considering the paraxial regime, as long as OAM calculations use a meaningful reference point. A discrepancy arrives, on the other hand, if the reference point is simply the origin. 
%(which corresponds to a type of geometrical center of the STVP)

%TTTTTTTTTTTTTTTTTTTTTTTTTTTTTTTTTTTTTTTTTTTTTTTTTTTTTTTTTT
\begin{table}[h]
\begin{center}
\caption{Centroids and corresponding OAM for the $xz$-elliptic and $xt$-elliptic STVPs in the uniform-mass model.}
\renewcommand{\arraystretch}{2}
\begin{tabular}{c| c | c  } 
\hline \hline 
 \textbf{Quantity} & \textbf{$xz$ STVP} & \textbf{$xt$ STVP}\\ [0.5ex] 
 \hline
 $x_{\rm cent}$ &0 & $\dfrac{\Delta_x w_t}{2}$  \\ 
  $x_{\rm min}$ & $-\dfrac{\Delta_x w_z}{2}$ &0  \\
 \hline 
 $\langle L_y\rangle(x_{\rm cent})$ &  $mc\dfrac{\Delta_x w_z}{2}$ &$ mc\dfrac{\Delta_x w_t}{2}$ \\
 $\langle L_y\rangle(x_{\rm min})$ &  $ 0$ &0 \\
 \hline\hline
\end{tabular}
 \label{tab:results-paraxial}
\end{center}
\end{table}
%TTTTTTTTTTTTTTTTTTTTTTTTTTTTTTTTTTTTTTTTTTTTTTTTTTTTTTTTTT

In the next section, we refine these results by leveraging a ``wavization'' of the particle model, to express them as a function of the STVP's topological charge $\ell$.

%%%%%%%%%%%%%%%%%%%%%%%%%%%%%%%%%%%%%%
\subsection{Quantization of vorticity}
\label{discretization}
%%%%%%%%%%%%%%%%%%%%%%%%%%%%%%%%%%%%%%
In order to relate the obtained OAM to a topological charge, one must perform a semi-classical quantization based on assigning phases along the rays according to their optical path length (OPL). The phase for each ray at the particle's position, referred to here as $\Phi(\xi)$, must be chosen to guarantee phase consistency between neighboring rays. This phase satisfies the following differential equation:
\begin{align}
\frac{\partial \Phi(\xi)}{\partial \xi} = k_0{\bf u}(\xi)\cdot\frac{\partial{\bf r}(\xi,t)}{\partial\xi}\,.
\label{eq:L-xi}
\end{align}
where $k_0 = mc$ is the wavenumber corresponding to the classical momentum of the particles. It is easy to show from Eqs.~\eqref{eq:generalrays} and \eqref{eq:px} that the right-hand side of this equation is independent of $t$. 
Without loss of generality, we choose $\Phi(0)=0$. Rigorous solutions to Eq.~(\ref{eq:L-xi}) are given in 
the Supplementary Document B
%Appendix~\ref{AppendixNew} 
for the particle loops in the $xz$ and $xt$ frameworks. Since the ray family is periodic, the ray construction is self-consistent from the wave-optical point of view  only if the values of $\Phi$ for $\xi=0$ and $\xi=2\pi$ (corresponding to the same ray) lead to consistent phases. 
That is, 
%if we define the phase as $\Phi(\xi)=k_0S(\xi)$, 
the condition $\Phi(2\pi)=2\pi\ell$ must be satisfied, where $\ell$ is an integer (corresponding to the STVP's topological charge). %and $\lambda_0$ is the wavelength. 
%As shown in the Supplementary Document B,
%Appendix~\ref{AppendixNew}, 
Since the underlying ray bundles are the same, this quantization leads to the same restriction for both frameworks, which in the paraxial regime takes the simple form
\begin{align}
k_0 \Delta_x w_x= 2 \ell\,.
\label{eq:paraxquant1}
\end{align}
%
%where $k_0=2\pi/\lambda_0$ is the wavenumber. 

Using the quantization relation \eqref{eq:paraxquant1}, the results of Table~\ref{tab:results-paraxial} can be expressed in terms of the topological charge $\ell$, as well as of the ellipses' semiaxis ratios $\gamma={w_z}/{w_x}$ (for $xz$ STVPs) and $\gamma = {w_t}/{w_x}$ (for $xt$ STVPs), 
%We also replace the mechanical total momentum $mc$ with the wave-optical one, $k_0$. 
as given in Table~\ref{tab:results-paraxial-l}. 
Surprisingly, these results with respect to either of the two centroids do not agree with any of the results for the intrinsic OAM in Table~\ref{tab:priorart}. 
Curiously, for $\gamma=1$, the OAM with respect to $x_{\rm cent}$ yields $\langle L_y \rangle =\ell$, the well-known formula for monochromatic spatial vortices \cite{Allen_book, Andrews_book, Bliokh2015PR}.
%We here rather find expressions for the orbital angular momentum of STOVs that are similar to those classically corresponding to the case of spatial vortices with longitudinal orbital angular momentum, with $\gamma_z$ and $\gamma_t$ playing the role of the vortex ellipticity $\gamma$. 
The origin of this discrepancy will be clarified in Section~\ref{sec:III}.

%TTTTTTTTTTTTTTTTTTTTTTTTTTTTTTTTTTTTTTTTTTTTTTTTTTTTTTTTTT
\begin{table}[h]
\begin{center}
\caption{OAM from Table~\ref{tab:results-paraxial} after the vortex quantization.}
\renewcommand{\arraystretch}{1.5}
\begin{tabular}{c| c | c  } 
 \hline \hline 
 \textbf{Quantity} & \textbf{$xz$ STVP} & \textbf{$xt$ STVP}\\ [0.5ex] 
 \hline
 $\langle L_y\rangle(x_{\rm cent})$ &  $  \gamma \ell$ &$ \gamma \ell$ \\
 $\langle L_y\rangle(x_{\rm min})$ &   0 &0 \\
 \hline\hline
\end{tabular}
 \label{tab:results-paraxial-l}
\end{center}
\end{table}
%TTTTTTTTTTTTTTTTTTTTTTTTTTTTTTTTTTTTTTTTTTTTTTTTTTTTTTTTTT
%\FloatBarrier

%FFFFFFFFFFFFFFFFFFFFFFFFFFFFFFFFFFFFFFFFFFFFFFFFFFFFFFFFFFFFFF
\begin{figure}[t]
\centering
\includegraphics[width=\linewidth]{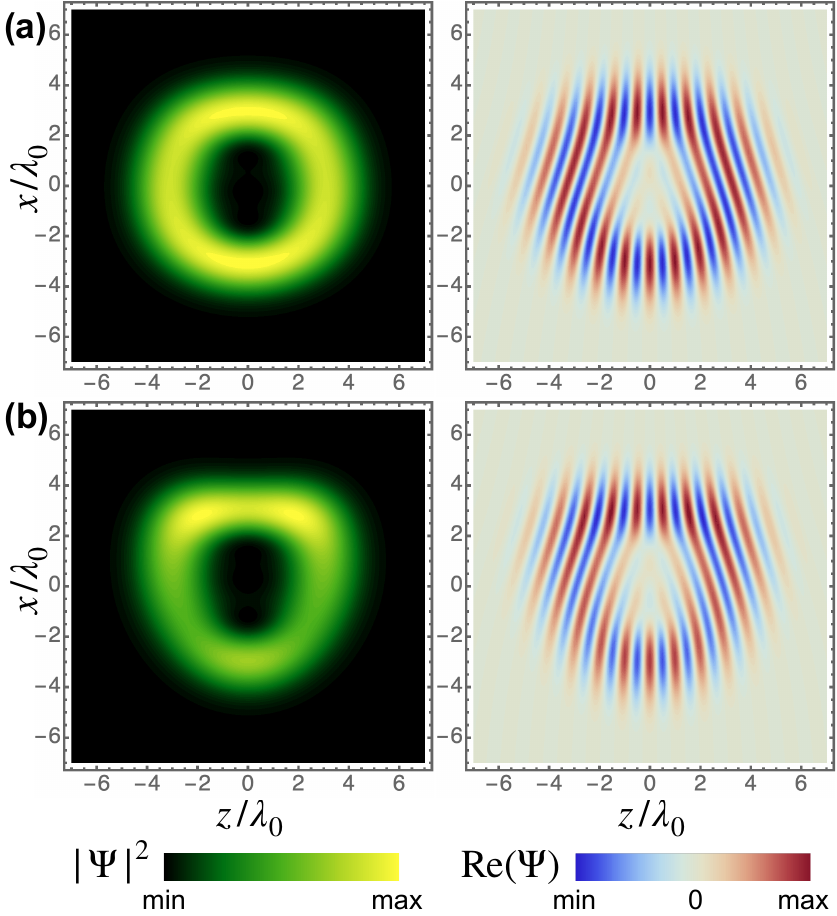}
\caption{Intensity and real part of the wavefield $\Psi(x,z,0)$, Eq.~\eqref{eq:field-safe}, for an $xz$ STVP (a) and an $xt$ STVP (b) with topological charge $\ell=3$, directional spread $\Delta_x=0.3$, and widths $w_x=w_z=w_t=3.29\lambda_0$ ($\gamma=1$), where $\lambda_0=2\pi/k_0$. The axes are in units of $\lambda_0$.}
\label{fig:STOV-fields-attempt1}
\end{figure}
%FFFFFFFFFFFFFFFFFFFFFFFFFFFFFFFFFFFFFFFFFFFFFFFFFFFFFFFFFFFFFF
%\FloatBarrier

%%%%%%%%%%%%%%%%%%%%%%%%%%%%%%%%%%%%%%
\subsection{Wave estimates through Gaussian wavepacket superposition}
\label{wavepacket1}
%%%%%%%%%%%%%%%%%%%%%%%%%%%%%%%%%%%%%%
One can estimate STVP wavefields from the particle-ray model by ``sewing the wave flesh on the classical bones'' \cite{Kravtsov1968, Berry1972}. For this, we dress each particle with a Gaussian wavepacket traveling in the direction of the ray. 
%Employing a quasi-monochromatic approximation with the central wavenumber $k_0$, 
The wave field at $t=0$ can be approximately written as
%
%\begin{widetext}
\begin{align}
&\Psi(x,z,0)=\!\int\limits_{0}^{2 \pi} e^{-\tfrac{[x-X(\xi,0)]^2+\gamma^{-2}[z-Z(\xi,0)]^2}{2 w_{\rm dress}^2}} \nonumber \\
&\times e^{\ui \Phi(\xi)+\ui k_0\left\{[x-X(\xi,0)]u_x(\xi)+ [x-Z(\xi,0)]u_z(\xi) \right\} } {\rm d}\xi\,,
%U(x,z,0)=\!\int\limits_{0}^{2 \pi} \exp\!\left\{-\frac{[x-X(\xi,0)]^2+[z-Z(\xi,0)]^2}{2 w_{\rm tot}^2}\right\} %\nonumber \\
%\!\exp\!\left\{\ui k_0 S(\xi)+\ui k_0 [x-X(\xi,0)]u_x(\xi)+ \ui k_0[x-Z(\xi,0)]u_z(\xi) \right\} {\rm d}\xi,
\label{eq:field-safe}
\end{align}
%\end{widetext}
%
where $w_{\rm dress}$ is the width of each Gaussian wavepacket,
chosen here as $w_{\rm dress}=(w_x/k_0\Delta_x)^{1/2}$. 
%The choice of the parameters $w_x$, $w_z$ (or $w_t$), $k_0$ and $\Delta_x$ allows tailoring the shape of the pulse. 
%Details of the method for choosing the Gaussians' width $w_{tot}$ in Eq.~\eqref{eq:field-safe} are given in Supplementary Document C.
%Appendix~\ref{AppendixB}. 
Figure~\ref{fig:STOV-fields-attempt1} shows the intensity and real part of an $xz$ STVP and an $xt$ STVP with topological charge $\ell=3$. Note that the global shapes of these intensity profiles are qualitatively consistent with those for the particle loops in Figs.~\ref{fig:Teq0_ZX}(c) and~\ref{fig:Zeq0_TX}(c), as well as with wave solutions \cite{porras2024clarification}. Note also that the wavefront spacing is fairly uniform over the whole pulse. 

\begin{figure}[t]
\centering
\includegraphics[width=\linewidth]{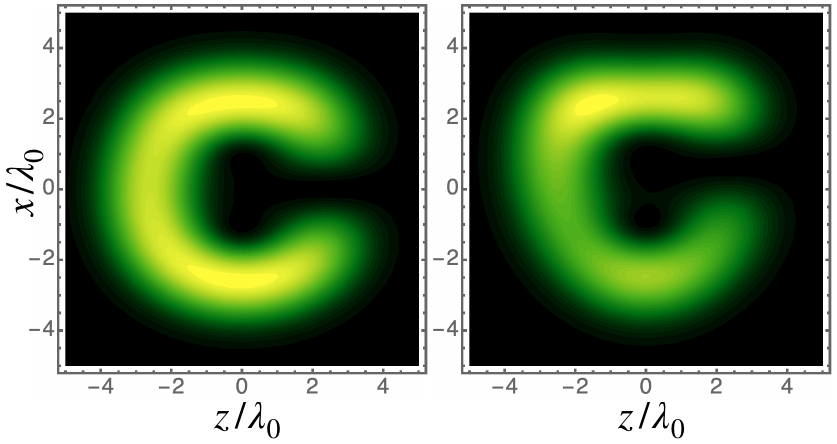}
\caption{Effect of the non-satisfaction of the quantization condition, $\Phi(2\pi) =  2\pi \ell$, on the wavefield intensity of an $xz$ STVP (left) and an $xt$ STVP (right), for $\Delta_x=0.3$ and  $w_x=w_z=w_t=2.8\lambda_0$, which give $\Phi(2\pi)/2\pi=2.64$. The positions of the interruptions depend on the chosen region of integration, in this case $\xi\in[0,2\pi)$.%\red{[KB: It is worth to explicitly indicate the value of $\Phi(2\pi)$ and parameters here.]}
}
\label{fig:nonsatis-selfconsistency}
\end{figure}

The quantization condition $\Phi(2\pi)=2\pi\ell$ guarantees that the integrand in \eqref{eq:field-safe} is periodic, so the integral is independent of the choice of limits as long as the interval covers the complete loop. 
If we choose particle loop dimensions for which $\Phi(2 \pi)$ is not an integer multiple of $2 \pi$, on the other hand, the STVP intensity profile does not close properly due to a phase inconsistency at the limits of integration, as shown on Fig.~\ref{fig:nonsatis-selfconsistency}. 
%\red{[KB: not clear about the pulse widths, can be confused with the Gaussian wavepacket width mentioned in this section. MA: Does the corrected wording clarify this?]}
Such wavefields can model {\it fractional} STVPs \cite{Huang2025}, analogous to the previously studied fractional spatial vortices \cite{Berry2004JOA, Leach2004NJP}.

%\textcolor{red}{Make connection to fractional stovs?}
%\begin{figure}[h!]
%\centering
%\includegraphics[width=120mm]{nonclosure.jpg}
%\caption{Effect of the non satisfaction of the self-consistency condition }
%\label{fig:nonsatis-selfconsistency}
%\end{figure}

%FFFFFFFFFFFFFFFFFFFFFFFFFFFFFFFFFFFFFFFFFFFFFFFFFFFFFFFFFFFFFF
%FFFFFFFFFFFFFFFFFFFFFFFFFFFFFFFFFFFFFFFFFFFFFFFFFFFFFFFFFFFFFF

%Similarly, the mismatch between the nonparaxial regime (Eqs.~(\ref{eq:Lxi-3Dspace}) and~(\ref{eq:Lxi-3Dspace-parax})), and the paraxial regime (Eqs.~(\ref{eq:Lxi-spacetime}) and~(\ref{eq:Lxi-spacetime-parax})) can be observed for a large value of $\Delta_x$. Indeed, if one considers a set of parameters satisfying the self-consistency condition for Eq.~(\ref{eq:Lxi-3Dspace}) or~(\ref{eq:Lxi-spacetime}), one can observe that this condition is not satisfied by the paraxial counterpart equations~(\ref{eq:Lxi-3Dspace-parax}) and~(\ref{eq:Lxi-spacetime-parax}), so that the intensity profile does not close anymore, as shown on Fig.~\ref{fig:nonsatis-selfconsistency}b for the $xz$ STOVs.
%and~\ref{fig:comp-parax-non-spacetime}.
%%%%%%%%%%%%%%%PARAXIAL SEPARATE XZ %%%%%%%%%%%
%\begin{figure}[h!]
%\centering
%\includegraphics[width=120mm]{comp-nonpara-parax-ZX.jpg}
%\caption{Comparison of nonparaxial and paraxial expressions of $S(\xi)$ - 3D space framework }
%\label{fig:comp-parax-non-3Dspace}
%\end{figure}
%%%%%%%%%%%%%%%%%%%%%%%%%%%%%%%%%%%%%
%\FloatBarrier
%\begin{figure}[h!]
%\centering
%\includegraphics[width=120mm]{comp-nonpara-parax-TX.jpg}
%\caption{Comparison of nonparaxial and paraxial expressions of $S(\xi)$ - spacetime framework }
%\label{fig:comp-parax-non-spacetime}
%\end{figure} \\ \\

%%%%%%%%%%%%%%%%%%%%%%%%%%%%%%%%%%%%%%
\section{Ray-particle model with a non-uniform mass density}
\label{sec:III}
%%%%%%%%%%%%%%%%%%%%%%%%%%%%%%%%%%%%%%

%%%%%%%%%%%%%%%%%%%%%%%%%%%%%%%%%%%%%%
\subsection{Mimicking wave momenta}
%%%%%%%%%%%%%%%%%%%%%%%%%%%%%%%%%%%%%%
%\subsection{Principle}
In order to understand the discrepancy between the results in Tables~\ref{tab:priorart} and \ref{tab:results-paraxial-l}, we consider the correspondence between rays and plane waves in the wave spectrum. In this manner the elliptical initial distribution of particles in the $(x,z)$ or $(x,t)$ plane corresponds to the elliptical distribution of the wave spectrum in the $(k_x,k_z)$ or $(k_x,\omega)$ plane \cite{bliokh2023orbital}. Importantly, these distributions involve plane waves with different frequencies and wavenumbers, i.e., {\it energies and momentum magnitudes}, while the simple ray-mechanical model of Section~\ref{sec:II} assumes particles with equal energies and momentum magnitudes.
This discrepancy can be resolved by introducing a nonuniform mass distribution for the classical particles, which modifies their energy and momentum-magnitude distributions in a proportional manner, similar to frequencies and wavenumbers of relativistic particles (photons). We therefore consider a ray family with momentum distribution ${\bf p}(\xi)$ mimicking the elliptical wavevector distribution in \cite{bliokh2023orbital}: 
\begin{align}
{\bf p}(\xi)=
\frac{m c}{2\pi} \left(\begin{array}{c}\Delta_x \cos{\xi}\\
1 - \Delta_z \sin{\xi}\end{array}\right).
\label{eq:p}
\end{align}
In the paraxial regime, $(\Delta_x,\Delta_z ) \ll 1$.
That is, keeping the particles' speed $c$ as constant, we have a mass density proportional to $|{\bf p}(\xi)|$.
Note that the corresponding unit vectors slightly differ from those in Eq.~\eqref{eq:px} but coincide with them in the paraxial regime:
\begin{align}
{\bf u}(\xi)=
\frac{{\bf p}(\xi)}{|{\bf p}(\xi)|} \simeq \left(\begin{array}{c}\Delta_x \cos{\xi}\\1\end{array}\right).
\label{eq:u_p}
\end{align}
%The paraxial-approximation formulas can be obtained by expanding Eqs.~\eqref{eq:p} and \eqref{eq:u_p} up to quadratic order in $\Delta_x$ and $\Delta_z$.

\begin{figure}[t]
     \centering
     \includegraphics[width=\linewidth]{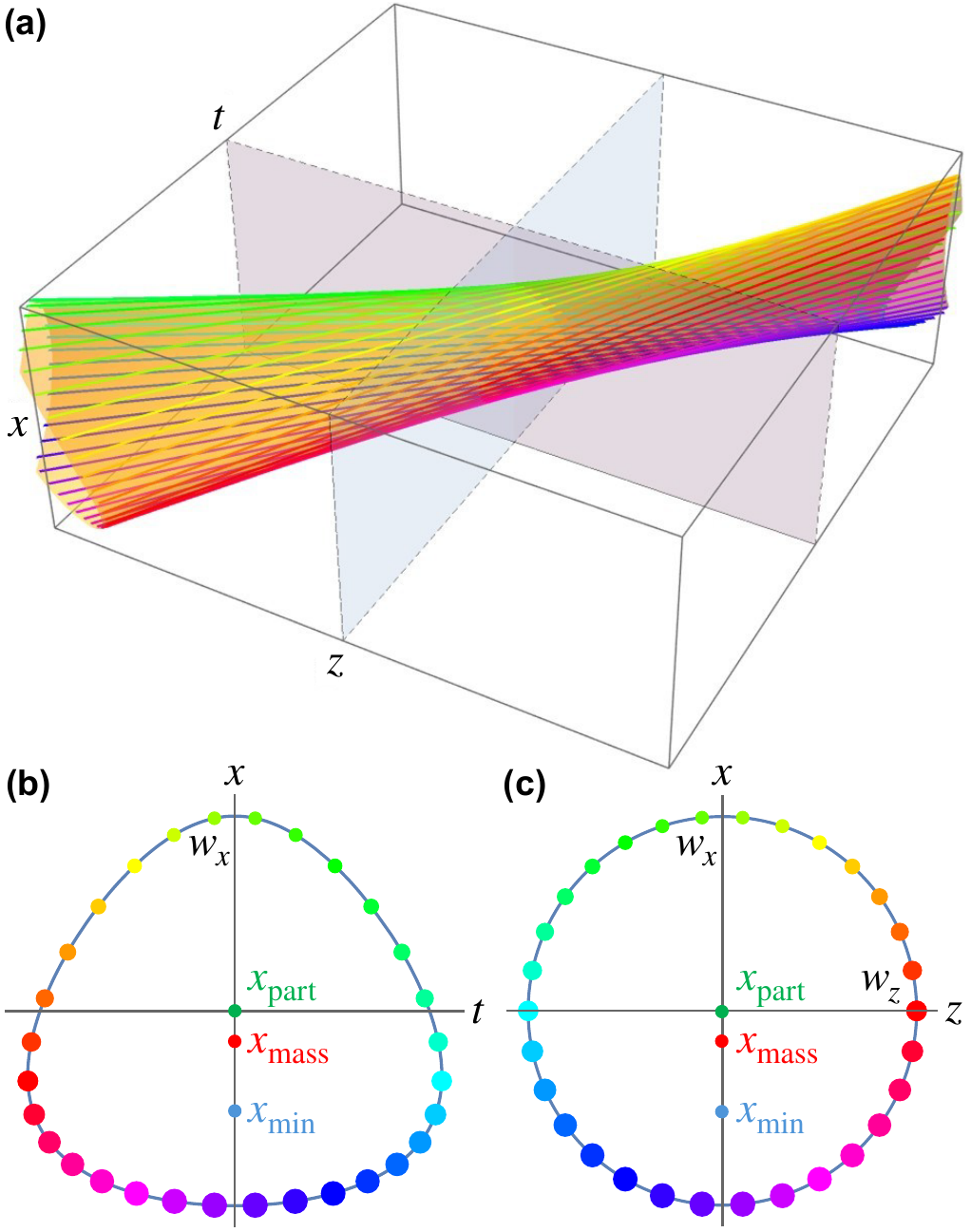}
     \caption{Ray-optics modeling of $xz$ STVPs with non-uniform particle mass density (indicated by the particles' sizes): (a) Ray bundle in space and time with $w_x=w_z$ and $\Delta_x= \Delta_z=0.3$; (b,c) Cross-sections at $z=0$ (b), and $t=0$ (c), where the reference points $x_{\rm mass}$, $x_{\rm part}$ and $x_{\rm min}$ are also indicated.}
      \label{fig:Teq0_ZXNEW}
\end{figure}

\begin{figure}[t]
\centering
\includegraphics[width=\linewidth]{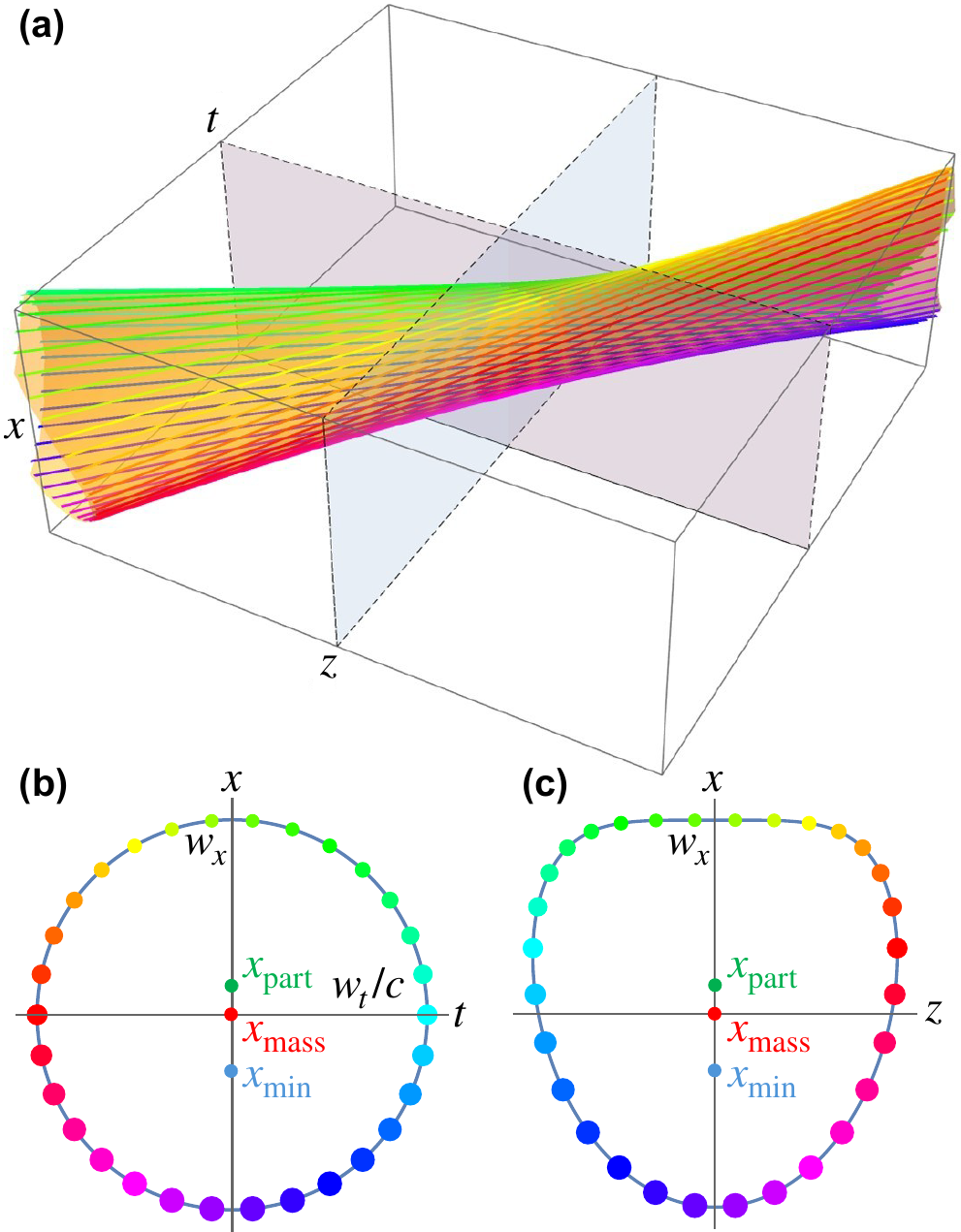}
\caption{ Ray-optics modeling of $xt$ STVPs with non-uniform particle mass density (indicated by the particles' sizes): (a) Ray bundle in space and time with $w_x=w_t$ and $\Delta_x=\Delta_z=0.3$; (b,c) Cross-sections at $z=0$ (b), and $t=0$ (c), where the reference points $x_{\rm mass}$, $x_{\rm part}$ and $x_{\rm min}$ are also indicated.}
\label{fig:Zeq0_TXNEW}
\end{figure}

We keep the same initial conditions as in Section~\ref{subsection:frameworks}: Eqs.~\eqref{Initial_1} for $xz$ STVPs and \eqref{Initial_2} for $xt$ STVPs.
Figures~\ref{fig:Teq0_ZXNEW} and \ref{fig:Zeq0_TXNEW}  are then the counterparts of Figs.~\ref{fig:Teq0_ZX} and \ref{fig:Zeq0_TX} with the slightly modified ray directions in Eqs.~\eqref{eq:p} and \eqref{eq:u_p}. 
Supplementary Movies 3 and 4 illustrate the corresponding propagation of the particle loops.

%FFFFFFFFFFFFFFFFFFFFFFFFFFFFFFFFFFFFFFFFFFFFFFFFFFFFFFFFFFFFFF
%FFFFFFFFFFFFFFFFFFFFFFFFFFFFFFFFFFFFFFFFFFFFFFFFFFFFFFFFFFFFFF

Two observations can be made. First, the non-elliptical shapes in Figs.~\ref{fig:Teq0_ZXNEW}(b) and \ref{fig:Zeq0_TXNEW}(c) differ slightly from the non-elliptical shapes in Figs.~\ref{fig:Teq0_ZX}(b) and \ref{fig:Zeq0_TX}(c), while still being similar to the wave-intensity shapes in \cite{porras2024clarification}. Second, the dot size in Figs.~\ref{fig:Teq0_ZXNEW} and \ref{fig:Zeq0_TXNEW} is proportional to the corresponding non-uniform mass density. This varying mass density causes a shift of the effective mass/energy centroid. 
%\red{[MA: I removed "away from the origin" because in one case it brings it closer to it.]}

%FFFFFFFFFFFFFFFFFFFFFFFFFFFFFFFFFFFFFFFFFFFFFFFFFFFFFFFFFFFFFF
%FFFFFFFFFFFFFFFFFFFFFFFFFFFFFFFFFFFFFFFFFFFFFFFFFFFFFFFFFFFFFF

%%%%%%%%%%%%%%%%%%%%%%%%%%%%%%%%%%%%%%
\subsection{Orbital angular momentum}
%%%%%%%%%%%%%%%%%%%%%%%%%%%%%%%%%%%%%%
We compare again the $xz$ and $xt$ STVP frameworks by computing their OAM. 
%which is defined in this case by 
%
%\begin{align}
%\langle{\bf L}\rangle(x_0)= \int\limits_{0}^{2 \pi}[{\bf r}(\xi,t) - {\bf r}_0] \times {\bf p}(\xi)\,{\rm d}\xi\,.
%\label{eq:OAMNEW}
%\end{align}
%\textcolor{blue}{(At some point we will have to comment on units. The momentum for the particles is dimensionless, since we are ignoring their ``mass'' and not mentioning their velocity ($c$). When we ``waveize'' we simply multiply by the average wavenumber $k_0$.)}
%Here, 
%${\bf r}(\xi,t)$ is determined by Eq.~\eqref{eq:generalrays} with ${\bf u} \to {\bf u}_{\bf p}$ (so that 
%$\langle{\bf L}\rangle$ 
%
%is time-independent (for a time-independent reference point $x_0$ and can be calculated at $t=0$, whereas the non-normalized vector ${\bf p}(\xi)$ intervenes in the definition of the OAM to account for the non-uniform mass density.
Substituting Eqs.~\eqref{eq:generalrays}--\eqref{Initial_2} and \eqref{eq:p} into Eq.~\eqref{eq:OAM}, we find that the OAM in both the $xz$ and $xt$ frameworks are given in the paraxial regime by [cf. Eqs.~\eqref{eq:OAM_1}]:
%As the OAM is computed at $t=0$, the results in the nonparaxial  and paraxial regimes will be identical for an $xz$-elliptical STVP 
%%%%%%%%%%%%%%%%%%%%%%%%%%%%%%%%%%%%%%%%%%%%%  NEW VERSION %%%%%%%%%%%%%%%%%%%%%%%%%%%%%%
%for which Eq.~(\ref{eq:OAMNEW}) yields
%
\begin{align}
%\label{eq:OAM_3Dspace_nonuniform}
\langle L_y^{(xz)}\rangle(x_0)
%&= - \frac{1}{ 2 \pi} \left[ \pi \Delta \left( w_x + \gamma w_z \right)+ 2 \pi x_0 \right]\nonumber \\
& = m c \left(x_0 +  \frac{\Delta_z w_x + \Delta_x w_z}{2} \right), \nonumber \\
\langle L_y^{(xt)} \rangle(x_0) & =  m c \left(x_0 + \frac{\Delta_z w_x}{2} \right).
\label{eq:OAM_ST_nonuniform}
\end{align}
That is, in both frameworks the non-uniformity of the particles' mass density (proportional to $\Delta_z$) causes an increase by $mc\Delta_zw_x/2$.
Similarly to Section~\ref{sec:II}, these new OAM expressions can be computed for different centroids $x_0$. While the centroid that minimizes the squared OAM, $x_{\rm min}$, is still defined by Eq.~(\ref{eq:OAM-min}), the particle-beam centroid can now be defined in two ways:
\begin{itemize}
    \item First, one can calculate the {\it mass centroid} (which corresponds to the wavepacket's {\it energy centroid}), which takes into account their nonuniform $\xi$-distributions:
%note that Eq.~(\ref{eq:xcent}) is no longer valid for the mass centroid, since the non-uniform mass density must be taken into account; the expression for $x_{\rm cent}$ is now given by
%
\begin{align}
x_{\rm mass} = \frac{1}{mc} \int\limits_{0}^{2 \pi} x(\xi,t) \left|{\bf p}(\xi)\right|\, {\rm d}\xi\,.
\label{eq:x0phot}
\end{align}
It is easy to show that this expression is time-independent, and one can use $\left|{\bf p}(\xi)\right| \simeq (mc/2\pi)(1-\Delta_z \sin\xi)$ in the paraxial approximation.

\item Second, one can use the geometric {\it particle centroid}, weighted not by mass/energy density but by particle number density \cite{bliokh2023orbital, Bliokh2012PRL}. This centroid is defined similarly to Eq.~\eqref{eq:xcent}:
\begin{align}
x_{\rm part}=\frac{1}{2 \pi} \int\limits_{0}^{2 \pi}x(\xi,t)\, {\rm d}\xi\,.
\label{eq:xphot}
\end{align}
\end{itemize}
%
%Of course, for uniform mass density $x_{\rm part}$ would coincide with $x_{\rm cent}$. 
The centroid \eqref{eq:xphot} is time-independent in our model. 
%\red{[MA: I checked this. In theory this could have been true since this centroid moves in the direction of the average ${\bf u}$ and not ${\bf p}$, but it turns out that, for our equations, the average of both $u_x$ and $p_x$ is zero. Can we can reword this?]} 
%\red{In the nonparaxial case, however, the particle centroid ${\bf r}_{\rm part}(t)$ propagates along a straight line non-collinear with the mean momentum (i.e., the $z$-axis in our case), which may result in non-conservation of the OAM defined with respect to such time-varying centroid $x_{\rm part}(t)$ \cite{porras2024clarification, Hancock2024PRX, Bliokh2025PLA}}. 
Note, however, that in nonparaxial wave solutions \cite{porras2024clarification, Hancock2024PRX, Bliokh2025PLA} the particle centroid ${\bf r}_{\rm part}(t)$ can propagate along a straight line that is non-collinear with the mean momentum, which would result in non-conservation of the OAM defined with respect to such time-varying centroid $x_{\rm part}(t)$. 
%However, for the model used here both the mean momentum and the trajectory traced by ${\bf r}_{\rm part}(t)$ coincide with the $z$ direction.

Table~\ref{tab:results-OAM-all} summarizes the paraxial results for the OAM for the three different reference points in the $xz$ and $xt$ frameworks.
Again, we note that these paraxial OAM calculations are independent of the particular model ($xz$ or $xt$) used for the particle loop. We also observe that the result for $\langle L_y\rangle(x_{\rm mass})$ is fully consistent with $\langle L_y\rangle(x_{\rm cent})$ in Table~\ref{tab:results-paraxial}, while $\langle L_y\rangle(x_{\rm part})$ picks up an extra term proportional to $\Delta_z$. On the other hand, the use of a nonuniform mass distribution does cause a change in the result for $\langle L_y\rangle(x_{\rm min})$ with respect to that in Table~\ref{tab:results-paraxial}. However, all results in Table~\ref{tab:results-OAM-all} are fully consistent with those in Table~\ref{tab:results-paraxial} in the limit $\Delta_z\to0$.

%
%%%%%%%%%%%%%%%% OLD RESULTS %%%%%%%%%%%%%
%\begin{table}[h]
%\begin{center}
%\caption{STOV Orbital angular momentum - non-uniform particle density \textcolor{blue}{CHECK RESULT FOR XMIN}}
%\begin{tabular}{c| c | c   } 
% \hline \hline 
% \textbf{Quantity} & \textbf{$xz$ STOV} & \textbf{$xt$ STOV paraxial} \\ [0.5ex] 
% \hline
% $x_{\rm cent}$ &0 & $\frac{ \gamma^2 \Delta w_x}{2}$   \\ 
%  $x_{0\rm min}$ & $-\Delta w_x$  & $ - \Delta w_x $   \\
%   $x_{\rm phot}$ & $-\frac{ \Delta w_x }{ 2 }$  & $ - \frac{  \Delta w_x}{2} \left( 1 - \frac{ \gamma^2}{2}  \right) $   \\
% $\langle L_y\rangle(x_{\rm cent})$ &  $ mc \frac{\Delta w_x}{2}  \left( 1 + \gamma^2 \right)$  & $ mc \frac{\Delta w_x}{2} ( 1 + \gamma^2) $  \\
% $\langle L_y\rangle(x_{0\rm min})$ &  $ - mc \frac{ \Delta w_x }{ 2}$  & $- mc \frac{\Delta w_x}{2}$ \\
%  $\langle L_y\rangle(x_{\rm phot})$ &  $ mc\frac{ \gamma^2 \Delta w_x}{2}$  & $ mc \frac{ \gamma^2 \Delta w_x}{2}  $
%   \\
% \hline
%\end{tabular}
% \label{tab:results-OAM-all}
%\end{center}
%\end{table}
%%%%%%%%%%%%%%%%%%%%%%%%%%%%%%%%%%%%%%%%%%
%
%TTTTTTTTTTTTTTTTTTTTTTTTTTTTTTTTTTTTTTTTTTTTTTTTTTTTTTTTTT
\begin{table}[h]
\begin{center}
\caption{Centroids and corresponding OAM for the $xz$-elliptic and $xt$-elliptic STVPs in the nonuniform-mass model.}
\renewcommand{\arraystretch}{2}
\begin{tabular}{c| c | c   } 
 \hline \hline 
 \textbf{Quantity} & \textbf{$xz$ STVP} & \textbf{$xt$ STVP} \\ [0.5ex] 
 \hline
    $x_{\rm mass}$ & $-\dfrac{ \Delta_z w_x }{ 2}$  & $ \dfrac{  \Delta_x w_t - \Delta_z w_x}{2}$   \\
 $x_{\rm part}$ &0 & $\dfrac{ \Delta_x w_t}{2}$   \\ 
   $x_{\rm min}$ & 
   $-\Delta_zw_x-\dfrac{\Delta_xw_z}2$
   %\red{$-\Delta_x w_z \dfrac{2+ \gamma^2}{2\gamma^2}$}  
   & 
   $-\Delta_zw_x$
   %\red{$- \dfrac{\Delta_x w_t}{\gamma^2}$} 
   \\
 \hline
$\langle L_y\rangle(x_{\rm mass})$ &  $ mc\dfrac{ \Delta_x w_z}{2}$  & $ mc \dfrac{ \Delta_x w_t}{2}  $
   \\
$\langle L_y\rangle(x_{\rm part})$ &  $ mc \dfrac{\Delta_z w_x + \Delta_x w_z}{2}$  & $ mc \dfrac{\Delta_z w_x+ \Delta_x w_t}{2}$  \\
 $\langle L_y\rangle(x_{\rm min})$ &  
 $-mc\dfrac{\Delta_zw_x}2$
 %\red{$- mc \dfrac{ \Delta_x w_z }{ 2\gamma^2}$} 
 & 
 $-mc\dfrac{\Delta_zw_x}2$
 %\red{$- mc \dfrac{\Delta_x w_t}{2\gamma^2}$} 
 \\
\hline\hline
\end{tabular}
 \label{tab:results-OAM-all}
\end{center}
\end{table}
%TTTTTTTTTTTTTTTTTTTTTTTTTTTTTTTTTTTTTTTTTTTTTTTTTTTTTTTTTT
%\FloatBarrier
%
%The results corresponding to non-uniform particle mass density are the same as previously as regards the independence of the orbital angular momentum on the STOV shape, for all considered centroids. Similarly to the previous calculations with uniform particle mass density, we will leverage the wavization of the particle model so as to express the new expressions as a function of the STOV topological charge $\ell$. 

%%%%%%%%%%%%%%%%%%%%%%%%%%%%%%%%%%%%%%
\subsection{Quantization of vorticity}
%%%%%%%%%%%%%%%%%%%%%%%%%%% NEW VERSION %%%%%%%%%%%%%%%%%%%%%%%%%%%%%%ùù
We again can express the OAM in terms of the topological charge $\ell$ by discretizing the phase accumulated over the loop. However, the use of a non-uniform momentum density translates in the wave domain to using a wavevector ${\bf k}(\xi)=k_0 ({2\pi}/{mc}) {\bf p}(\xi) = 2\pi {\bf p}(\xi)$ , 
%\red{[MA: Should we remove the $2\pi$?] \blue{[KB: I think it is needed for the momentum definition (11).]}}, 
whose non-uniform magnitude is centered at $k_0$. We then consider a phase $\Phi$ defined similar to Eq.~\eqref{eq:L-xi}:
\begin{align}
\frac{\partial \Phi(\xi)}{\partial \xi} = 
{\bf k}(\xi)\cdot\frac{\partial {\bf r}(\xi,t)}{\partial\xi}\,.
\label{eq:L-xi-new-slopes1}
\end{align}
Integrating Eq.~(\ref{eq:L-xi-new-slopes1}) in the paraxial approximation and imposing the quantization condition $\Phi(2\pi) = 2\pi\ell$, we obtain similar results in the $xz$ and $xt$ frameworks: 
%
%\begin{align}
%\Phi(\xi)= k_0 \gamma w_x \left(\Delta_z\xi + \cos{\xi} - 1   \right),
%\end{align}
%
%which satisfies $\Phi(0) = 0$. %\textcolor{red}{check results... they seem to differ by a factor of 2 from those of previous section.}
%Enforcing consistency of phase as the loop is closed implies the condition $\Phi(2\pi)=2\pi\ell$ with integer $\ell$, which results in
\begin{align}
k_0 ({\Delta_x w_x + \Delta_z w_{z,t}}) = 2\ell\,.
\label{eq:paraxquantvaryingdensity}
\end{align}
%\begin{align}
% \Phi(2 \pi)=2 \pi k_0 \gamma \Delta_z w_x=2 \pi k_0  \Delta_x w_x.   
%\end{align}
%\end{table}
%\red{[MA: If we delay the wave-based assumption for the gammas, this becomes $$k_0\frac{\Delta_xw_x+\Delta_zw_z}2=\ell,$$ which we can write as
%$$k_0\frac{\Delta_xw_x(1+\gamma/\bar{\gamma})}2=\ell,$$
%so we can see that the ratio of the gammas changes the quantization. For Section II this ratio is zero, while with the careful wave criterion it is 1.]}
This quantization condition differs from that in Eq.~\eqref{eq:paraxquant1} through the appearance of a term proportional to $\Delta_z$, related to the non-uniform weight of the particles. 

Similarly to Section~\ref{sec:II}, we introduce the ratio of the STVP ellipse semiaxes in real space, $\gamma = {w_z}/{w_x} = w_t/w_x$, which should also correspond to the inverted relation for the momentum-space ellipse in Eq.~\eqref{eq:p}: $\Delta_z/\Delta_x = 1/\gamma$. Using this parameter and the quantization condition \eqref{eq:paraxquantvaryingdensity}, we re-write the OAM from Table~\ref{tab:results-OAM-all} in Table~\ref{tab:results-OAM-NEW-with l21}.
%Using the quantization condition \eqref{eq:paraxquantvaryingdensity}, the OAM results in Table~\ref{tab:results-OAM-all} can be expressed in terms of the topological charge $\ell$, as shown in Table~\ref{tab:results-OAM-NEW-with l21}.  

Remarkably, the resulting expressions precisely reproduce the results obtained in \cite{hancock2021mode, bliokh2023orbital, bekshaev2024spatiotemporal, Porras2023OL} and shown in Table~\ref{tab:priorart}. 
First, the OAM with respect to the mass centroid coincides with $L_y = \gamma\ell/2$ in \cite{hancock2021mode, bliokh2023orbital, bekshaev2024spatiotemporal, Hancock2024PRX} and $L^{\rm INT}_y = \gamma\ell/2$ in \cite{porras2023transverse, porras2024clarification} (all these results were calculated with respect to the electromagnetic-energy centroid). This OAM is half-integer for a circular STVP with $\gamma=1$. Second, the OAM with respect to the particle centroid coincides with $L^{\rm INT}_y=(\gamma+\gamma^{-1})\ell/2$ calculated in \cite{bliokh2023orbital, bekshaev2024spatiotemporal} with respect to the `photon-probability' centroid. This OAM becomes integer for $\gamma=1$ and has the same form as longitudinal OAM of monochromatic vortex beams.

%TTTTTTTTTTTTTTTTTTTTTTTTTTTTTTTTTTTTTTTTTTTTTTTTTTTTTTTTTT
\begin{table}[h]
\begin{center}
\caption{OAM from Table~\ref{tab:results-OAM-all} after the vortex quantization.}
\renewcommand{\arraystretch}{2}
\begin{tabular}{c| c | c   } 
 \hline \hline 
 \textbf{Quantity} & \textbf{$xz$ STVP} & \textbf{$xt$ STVP} \\ [0.5ex] 
 \hline
  $\langle L_y\rangle(x_{\rm mass})$  & $\dfrac{\gamma}{2}\, \ell$ &  $\dfrac{\gamma}{2}\, \ell$ \\
 $\langle L_y\rangle(x_{\rm part})$ & $\dfrac{\gamma + \gamma^{-1}}{2}\ell$ & $\dfrac{\gamma + \gamma^{-1}}{2}\ell$  \\
 $\langle L_y\rangle(x_{\rm min})$ &   $-\dfrac{1}{2\gamma} \ell$      &  $-\dfrac{1}{2 \gamma} \ell$  \\
 \hline\hline
\end{tabular}
 \label{tab:results-OAM-NEW-with l21}
\end{center}
\end{table}
%TTTTTTTTTTTTTTTTTTTTTTTTTTTTTTTTTTTTTTTTTTTTTTTTTTTTTTTTTT
%\FloatBarrier

%%%%%%%%%%%%%%%%%%%%%%%%%%%%%%%%%%%%%%
\subsection{Wave estimates}
\begin{figure}[t!]
%\centering
\includegraphics[width=\linewidth]{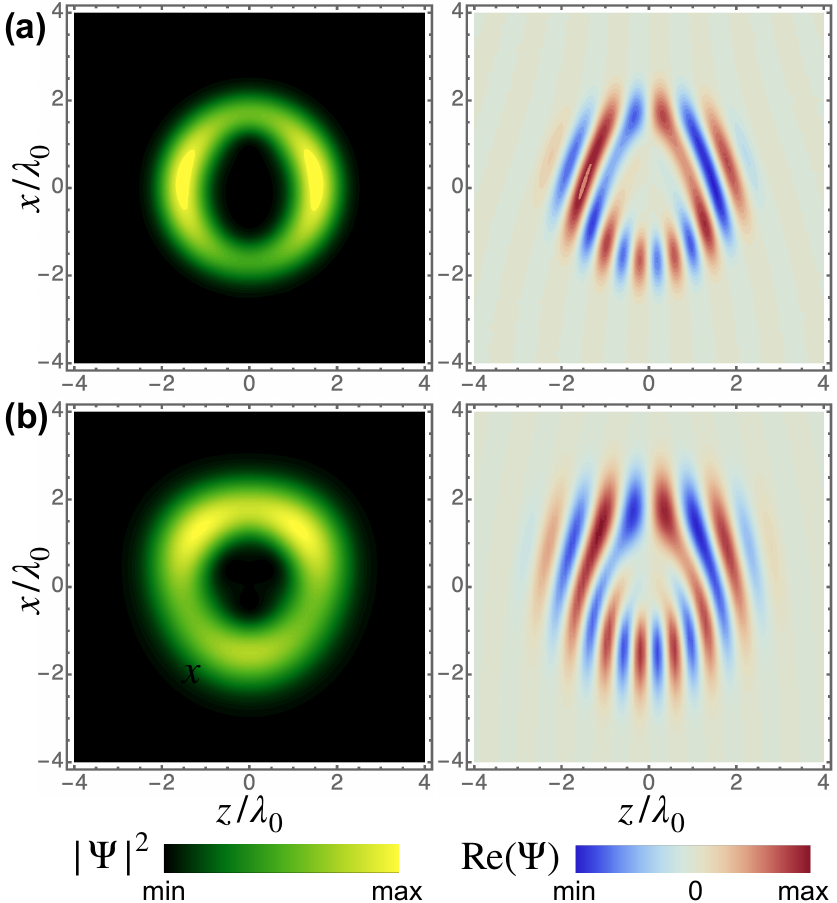}
\caption{Same as in Fig.~\ref{fig:STOV-fields-attempt1} but for the model with a non-uniform mass distribution. The parameters are $\ell=3$ and $w_x=w_z=w_t=1.59\lambda_0$.
%Intensity and real part of an $xz$ STVP and a $xt$ STVP with topological charge 3 in the ($x$,$z$) plane, with $k_0 w_x=k_0 w_z=k_0 w_t=10$, $k_0 w_{tot}=3.1$. The axes are in units of $\lambda_0=2\pi/k_0$.
}
\label{fig:STOV-l-3-newslopes}
\end{figure}

%%%%%%%%%%%%%%%%%%%%%%%%%%%%%%%%%%%%%%
Wave estimates can be calculated by using \eqref{eq:field-safe} with the substitution $k_0{\bf u}(\xi) \to {\bf k}(\xi) = 2\pi {\bf p}(\xi)$.
Figure~\ref{fig:STOV-l-3-newslopes} 
shows the intensity and real part of, respectively, an $xz$ STVP and an $xt$ STVP with topological charge $\ell=3$. 
Note that the global shapes of the intensity profiles in these figures are consistent with the ray loops of Figs.~\ref{fig:Teq0_ZXNEW}(c) and~\ref{fig:Zeq0_TXNEW}(c). In addition, differently from the case for uniform mass density, the plots for the real part now show different wavefront spacings at the top and bottom of the STVP, reflecting the fact that local wavelength is inversely proportional to mass weight. 

%FFFFFFFFFFFFFFFFFFFFFFFFFFFFFFFFFFFFFFFFFFFFFFFFFFFFFFFFFFFFFF
%FFFFFFFFFFFFFFFFFFFFFFFFFFFFFFFFFFFFFFFFFFFFFFFFFFFFFFFFFFFFFF
%\FloatBarrier

%%%%%%%%%%%%%%%%%%%%%%%%%%%%%%%%%%%%%%
\subsection{The role of dispersion}
\label{sec:dispersion}
%%%%%%%%%%%%%%%%%%%%%%%%%%%%%%%%%%%%%%

The OAM with respect to the mass centroid \eqref{eq:x0phot}, $\langle L_y\rangle(x_{\rm mass}) = \gamma\ell/2$ corresponds to the OAM of optical STVPs calculated with respect to the energy centroid \cite{hancock2021mode, bliokh2023orbital, bekshaev2024spatiotemporal, Hancock2024PRX}. For optical waves, which correspond to massless relativistic particles (photons), the energy, mass, and the absolute value of the momentum are all proportional to each other. This is expressed by the linear dispersion relation $\omega = k c$ or $E=pc = mc^2$. Therefore, our model and the OAM expressions are also valid for other waves with linear dispersion, such as sound waves in a fluid or gas \cite{bliokh2023orbital}. 

For waves with a nonlinear dispersion relation, this model and OAM equations are no longer valid. For example, let us consider STVPs of nonrelativistic quantum particles. The relation between the energy, mass, and momentum becomes $E\simeq mc^2 + p^2/2m$, where the second term is a small correction to the main rest energy. For such STVPs, the momentum distribution is still given by Eq.~\eqref{eq:p}, but calculating the mass or energy centroid \eqref{eq:x0phot} we need to change $|{\bf p}(\xi)| \to mc/2\pi$. This results in the equivalence of the mass and particle centroids (as expected in nonrelativistic physics), $x_{\rm mass} = x_{\rm part}$, and a uniquely defined intrinsic OAM with respect to this center: $\langle L_y \rangle (x_{\rm mass}) = (\gamma+\gamma^{-1})\ell/2$ \cite{bliokh2023orbital}.

%%%%%%%%%%%%%%%%%%%%%%%%%%%%%%%%%%%%%%
\section{Concluding remarks}
%%%%%%%%%%%%%%%%%%%%%%%%%%%%%%%%%%%%%%
We have presented a simple model using rays associated with the rectilinear paths of traveling particles to understand the OAM of STVPs. This model illustrates, for example, how STVPs can be elliptical either in space or in spacetime, but not in both simultaneously \cite{porras2024clarification}. Deformations of the STVP in the $(x,z)$ or $(x,t)$ planes produce a shift of its centroid; however, once this shift is properly accounted for in the OAM calculations, the resulting OAM is independent of whether the STVP is elliptical in space or in spacetime. 

We also considered two different models for the particle mass-density distribution: a uniform distribution and a non-uniform one chosen such that the mechanical momentum distribution mimics the spatial Fourier spectrum of a STVP. These different mass distributions do not affect the mechanical OAM calculated with respect to the mass centroid, at least within the paraxial regime (see Tables~\ref{tab:results-paraxial} and \ref{tab:results-OAM-all}). The similarity of these mechanical results reflects the fact that, in all cases, the particle distribution undergoes similar evolutions in all cases, including an approximately quarter-cycle rotation (ignoring other deformations) between $t=0$ and $|t|\to \infty$, as can be observed from the four Supplemental Movies. 
%[KB: I thought that this is related to diffraction and the Gouy phase rather than to the OAM value.]} 
However, the choice of mass distribution does have an important effect in the semiclassical correspondence between particles and waves and in the resulting vortex quantization conditions, Eqs.~\eqref{eq:paraxquant1} and \eqref{eq:paraxquantvaryingdensity}. Namely, for a given directional spread $\Delta_x$, the spatial width $w_x$ needed to achieve a given vorticity $\ell$ must be twice as large for the uniform mass density as for the non-uniform one. This difference can be understood by looking at the real-part field plots in Figs.~\ref{fig:STOV-fields-attempt1} and \ref{fig:STOV-l-3-newslopes}: for the latter (varying mass density), the different wavefront spacings on the top and bottom of the STVP naturally help accommodate a different number of fringes on both sides, so a given vorticity is achievable with a smaller loop. %[KB: I am not sure I understand this explanation. The STVP ring sizes and $\ell$ are very similar in these figures. In Fig. 8, the top/bottom spacings are larger/smaller than in Fig. 4, so this looks like redistribution of spacings along the loop.]} 

Using the model with a non-uniform mass distribution and the corresponding quantization conditions, we were able to reproduce the results of previous wave-based calculations \cite{hancock2021mode, bliokh2023orbital, porras2023transverse, porras2024clarification, bekshaev2024spatiotemporal, Tripathi2025OE} (see Tables~\ref{tab:priorart} and \ref{tab:results-OAM-NEW-with l21}). This includes the differing values of the OAM computed with respect to the mass/energy centroid and the particle centroid. Notably, this distinction also depends on the wave dispersion and the associated relation between the momentum and energy of particles. In the present work, we employed a classical-particle model with a varying mass density to mimic the proportionality between momentum and energy characteristic of free-space optical (or sound) waves and relativistic massless particles. For other types of waves and dispersion relations, the model would need to be modified accordingly.  

%The agreement of the OAM for the varying density model in Section~\ref{sec:III} with prior wave results is due to the fact that, unlike the uniform density model in Section ~\ref{sec:II}, this particle model resembles the wave model not only in the position representation but also in the momentum one. 

%This geometric optics-based model was then connected to a wave model by dressing each particle with a Gaussian wavepacket traveling along the direction of its corresponding ray. Fields plots were computed from this wave model. Finally, we derived STVPs orbital angular momentum expressions, demonstrating that those are independent on the STVP shape. We made different assumptions regarding the particle mass density, which let us conclude on the quantization in half-integer units of the STVP orbital angular momentum, as in \cite{hancock2021mode}, when computed at the mass centroid, and retrieve the result of \cite{bliokh2023orbital} when considering a photon centroid.

\vspace{1cm}

\section*{Acknowledgements}
MAA acknowledges funding from the Agence Nationale de Recherche (ANR) through the project 3DPol, ANR-21-CE24-0014-01. KYB acknowledges support from Marie Sk\l{}odowska-Curie COFUND Programme of the European Commission (project HORIZON-MSCA-2022-COFUND-101126600-SmartBRAIN3), 
ENSEMBLE3 Project carried out within the International Research Agendas Programme (IRAP) of the Foundation for Polish Science co-financed by the European Union under the European Regional Development Fund (MAB/2020/14) and Teaming Horizon 2020 programme of the European Commission (GA. No. 857543), 
and Minister of Science and Higher Education ``Support for the activities of Centers of Excellence established in Poland under the Horizon 2020 program'' (contract MEiN/2023/DIR/3797). 

We thank Miguel A. Porras and Etienne Brasselet for useful discussions.
We would like to dedicate this work to the memory of Kevin J. Parker, whose enthusiasm for interesting wavepackets was contagious.

%\clearpage

\bibliography{refs}
%\printbibliography

\newpage

%\twocolumngrid

%\bibliography{refs.bib}

\onecolumngrid
\newpage
\appendix

%\clearpage
%\appendix

\section{Centroids and OAM expressions in the nonparaxial regime}
\label{AppendixA}

The following tables show the expressions for the centroids and OAM calculated with Eqs.~(6),~(8), and (9) from the main document. Here, $E$ and $K $ designate respectively the complete elliptic integrals of the second and first kinds, and $\Delta_x$ was replaced by $\tan{\eta}$, where $\eta$ designates the maximum ray angle in the ray family.

\vspace{0 cm}

\begin{table}[ht!]
    \centering
    \begin{center}
\caption{Nonparaxial regime for an $xz$ STVP}
\hspace*{-1.7 cm}
\begin{tabular}{c| c  } 
 \hline \hline 
 \textbf{Quantity} & \quad \\ [0.5ex] 
 \hline
 $x_{\rm cent}$& 0   \\ 
  $x_{0\rm min}$ &  $- w_z \tan\dfrac{\eta}2$   \\
 $\langle L_y\rangle(x_{\rm cent})$ &  $-2 w_z \Big[ \dfrac{1}{\tan{\eta}} E(-\tan^2{\eta}) - 
   \dfrac{1}{\tan{\eta}} K(-\tan^2{\eta}) + $ \\
   \quad \quad \quad & $\left(\dfrac{1}{\sin{\eta}} E(\sin^2{\eta}) - 
   \cos^2{\eta}  K(\sin^2{\eta}) \right) \Big]$ \\
 $\langle L_y\rangle(x_{0\rm min})$ & $-2 w_z \Big[ \dfrac{1}{\tan{\eta}} E(-\tan^2{\eta}) +
   (\dfrac{2 }{\tan{\eta}} - \dfrac{1}{ \sin{\eta}}) K(-\tan^2{\eta})$ \\
 \quad  &  $+  \left(\dfrac{1}{\sin{\eta}} E(\sin^2{\eta}) + 
    \cos{\eta}(\dfrac{2 }{\tan{\eta}} - \dfrac{1}{ \sin{\eta}}) K(\sin^2{\eta}) \right) \Big]$  \\
 \hline
\end{tabular}
 \label{tab:results-nonparaxial-XZ2}
\end{center}
%\end{table} \\
\vspace{0 cm}
%\begin{table}[H]

%\end{table} 
\vspace{0 cm}
%\begin{table}[H]
\begin{center}
\caption{Nonparaxial regime for an {$xt$ STVP}}
\begin{tabular}{c | c  } 
 \hline \hline 
 \textbf{Quantity} & \quad \\ [0.5ex] 
 \hline
 $x_{\rm cent}$ & $-\dfrac{2 w_t}{\tan{\eta}} \Big[E(-\tan^2{\eta}) - 
    K(-\tan^2{\eta}) +  \left(\dfrac{1}{\cos{\eta}} E(\sin^2{\eta}) - 
   \cos{\eta}  K(\sin^2{\eta}) \right) \Big]$  \\ 
  $x_{0\rm min}$  &0  \\
 $\langle L_y\rangle(x_{\rm cent})$  &$-2x_{\rm cent} \Big[K(-\tan^2{\eta}) + \cos{\eta}  K(\sin^2{\eta})  \Big]$ \\
 $\langle L_y\rangle(x_{0\rm min})$  & $0$ \\
 \hline
\end{tabular}
 \label{tab:results-nonparaxial-XT2}
\end{center}
\end{table}

%%%%%%%%%%%%%%%%%%%%%%%
\section{Solutions for the phase}
\label{AppendixNew} 
The phase $\Phi(\xi)$ is meant to ensure that all ray contributions are added in phase.
Equation~(10) from the main document can be integrated both for $xz$ and $xt$ STVPs. By choosing the condition $\Phi(0)=0$, we find for the $xz$ STVPs:
\begin{align}
    \Phi(\xi)&=
k_0w_x \Bigg[\frac{\sqrt{1+ \Delta_x^2}}{\Delta_x} E\!\left(\xi,\frac{\Delta_x^2}{1+\Delta_x^2}\right)- \frac{1}{\Delta_x \sqrt{1+ \Delta_x^2} } F\!\left(\xi,\frac{\Delta_x^2}{1+\Delta_x^2}\right)   \Bigg] \nonumber \\
&+ k_0w_z \log{\!\left( \frac{\sqrt{2} \Delta_x \cos{\xi}+\sqrt{2 + \Delta_x^2 + \Delta_x^2 \cos{2 \xi}}}{\sqrt{2}\Delta_x + \sqrt{2+2 \Delta_x^2}}   \right)}, 
        \label{eq:Lxi-3Dspace}
\end{align}\\
and for $xt$ STVPs:
\begin{align}
   %%%%%%%%%%%%%% OLD CONVENTION %%%%%%%%%%%%%ù
    % S(\xi)&=\frac{w_t \tau \sin{\xi}+ w_x [F(\xi,-\tau^2)-  E(\xi,-\tau^2)] }{\tau},
     %%%%%%%%%%%%%%%%%%%%%%%%
     \Phi(\xi)&= k_0w_t \left( \cos{\xi} -1  \right)\nonumber\\
     &+ k_0w_x \left[\frac{\sqrt{1 + \Delta_x^2}}{\Delta_x} E\!\left(\xi,\frac{\Delta_x^2}{1+\Delta_x^2}\right) - \frac{1}{\Delta_x \sqrt{1 + \Delta_x^2} } F\!\left(\xi,\frac{\Delta_x^2}{1+\Delta_x^2}\right)  \right],
      \label{eq:Lxi-spacetime}
\end{align}
where $F$ and $E$ are, respectively, the elliptic integrals of the first and second kinds.
As a consequence, the integrand of Eq.~(12) in the main document can be expressed in an analytical form in both cases. Notice that both equations~(\ref{eq:Lxi-3Dspace}) and~(\ref{eq:Lxi-spacetime}) define similar functions that only differ in one term. %This similarity can be appreciated in Fig.~\ref{fig:Lxi}.
      
The choice of the parameters $w_x$, $w_z$ (or $w_t$), and $\Delta_x$ allows tailoring the shape of the pulse. 
As mentioned in the main text, the integral in $\xi$ should be independent of the choice of limits as long as the interval covers the complete loop. This means that the integrand must be a periodic function of $\xi$, which is only true if $\Phi(2 \pi) =  2 \pi \ell$ %is an integer multiple of $2 \pi$, that is, 
%\begin{align}
%    \Phi(2 \pi) =  2 \pi \ell,
%    \label{eq:selfconsistency}
%\end{align}
where the integer $\ell$ corresponds to the STVP's topological charge. For both types of STVP, this leads to the same constraint:
\begin{align}
    \frac{4k_0w_x}{\Delta_x} \Bigg[ \sqrt{1+ \Delta_x^2} E\!\left(  \frac{\Delta_x^2}{1+\Delta_x^2}\right) - \frac{1}{\sqrt{1+ \Delta_x^2}} K\!\left(  \frac{\Delta_x^2}{1+\Delta_x^2}\right) \Bigg] = 2 \pi \ell,
\end{align}
where $K$ and $E$ are, respectively, the complete elliptic integral of the first kind and the complete elliptic integral.  

The paraxial regime can be studied by approximating the expressions of $\Phi(\xi)$ of Eqs.~(\ref{eq:Lxi-3Dspace}) and~(\ref{eq:Lxi-spacetime}). By an expansion in $\Delta_x$ up to the second order, we find for the $xz$ STVPs:
\begin{align}
  %%%%%%%%%% OLD CONVENTION %%%%%%%%%%%%%%%%%%%%ù
   % S(\xi)\approx w_z \sin{\xi} + \frac{\tau}{4} w_x (-2 \xi + \sin{2\xi}) - \frac{\tau^2}{6} w_z (\sin{\xi})^3,
    %%%%%%%%%%%%%%%%%%%%%%%%%%%%%%%%%%%%%%%%%%%%
    \Phi(\xi)&\simeq k_0w_z \left(\cos{\xi} -1 \right)
    + \frac{k_0\Delta_x}{4} \left[ w_z + 2 w_x \xi - 2 w_z \cos^2{\xi} + w_z \cos{\left( 2 \xi  \right)} + w_x \sin{\left( 2 \xi  \right)}    \right) \nonumber \\
   & + \frac{k_0w_z \Delta_x^2}{12} \left(2 - 3 \cos{\xi} + 4\cos^3{\xi} -3 \cos{\xi} \cos{\left( 2 \xi  \right)}  \right],
        \label{eq:Lxi-3Dspace-parax}
\end{align}
and for the $xt$ STVPs:
\begin{align}
    %%%%%%%%%% OLD CONVENTION %%%%%%%%%%%%%%%
    % S(\xi) \approx  w_t \sin{\xi} + \frac{\tau}{4} w_x(- 2\xi + \sin{2 \xi}).
     %%%%%%%%%%%%%%%%%%%%%%%%%%%%%%%%
     \Phi(\xi) \simeq  k_0w_t \left(\cos{\xi} -1 \right) +\frac{\Delta_x}{4} k_0w_x( 2\xi + \sin{2 \xi}).
      \label{eq:Lxi-spacetime-parax}
\end{align}
The constraint is again the same in both frameworks: 
\begin{align}
   \Phi(2 \pi)= k_0 \pi w_x \Delta_x= 2 \pi \ell.
   \label{eq:paraxquant1}
\end{align}

\end{document}